\documentclass[aps,prd,twocolumn,showpacs,superscriptaddress,groupedaddress,floatfix]{revtex4-1} 

\usepackage{graphicx, color}  % needed for figures
\usepackage{dcolumn}   % needed for some tables
\usepackage{bm}        % for math
\usepackage{amssymb}   % for math
\usepackage{amsmath}   % for math
\usepackage{footmisc}
\usepackage{enumerate}
\usepackage{comment}
\usepackage{tabularx}
\usepackage{booktabs}
\usepackage{placeins}

\usepackage{hyperref}
\usepackage{siunitx}
\usepackage{multirow} %Used to make the tables look nice in two-column format

\newcommand{\RN}[1]{%
  \textup{\uppercase\expandafter{\romannumeral#1}}%
}

\begin{document}

\preprint{APS/123-QED}

\title{Improving the stability of frequency dependent squeezing with bichromatic control of filter cavity length, alignment and incident beam pointing}

\author{Yuhang Zhao$^{1, 2}$} 
\email[Corresponding author:]{yuhang@icrr.u-tokyo.ac.jp}
\author{Eleonora Capocasa$^{3}$}
\author{Marc Eisenmann$^{2}$}
\author{Naoki Aritomi$^{2}$}
\author{Michael Page$^{2}$}
\author{Yuefan Guo$^{4}$} 
\author{Eleonora Polini$^{5}$}
\author{Koji Arai$^{6}$}
\author{Yoichi Aso$^{2}$}
\author{Martin van Beuzekom$^{4}$}
\author{Yao-Chin Huang$^{7}$} 
\author{Ray-Kuang Lee$^{7}$}
\author{Harald L{\"u}ck$^{8}$}
\author{Osamu Miyakawa$^{1}$} 
\author{Pierre Prat$^{3}$} 
\author{Ayaka Shoda$^{2}$} 
\author{Matteo Tacca$^{4}$} 
\author{Ryutaro Takahashi$^{2}$} 
\author{Henning Vahlbruch$^{8}$} 
\author{Marco Vardaro$^{4,9,10}$} 
\author{Chien-Ming Wu$^{7}$} 
\author{Matteo Leonardi$^{2}$}
\author{Matteo Barsuglia$^{3}$} 
\author{Raffaele Flaminio$^{5,2}$} 

\address{$^{1}$Institute for Cosmic Ray Research (ICRR), KAGRA Observatory, The University of Tokyo, Kamioka-cho, Hida City, Gifu 506-1205, Japan}
\address{$^{2}$National Astronomical Observatory of Japan, 2-21-1 Osawa, Mitaka, Tokyo, 181-8588, Japan}
\address{$^{3}$Laboratoire Astroparticule et Cosmologie (APC), 10 rue Alice Domon et L\'eonie Duquet, 75013 Paris, France}
\address{$^{4}$Nikhef, Science Park, 1098 XG Amsterdam, Netherlands}
\address{$^{5}$Laboratoire d'Annecy-le-Vieux de Physique des Particules (LAPP), Université Savoie Mont Blanc, CNRS/IN2P3, F-74941 Annecy-le-Vieux, France}
\address{$^{6}$LIGO, California Institute of Technology, Pasadena, California 91125, USA}
\address{$^{7}$Institute of Photonics Technologies, National Tsing-Hua University, Hsinchu 300, Taiwan}
\address{$^{8}$Institut f{\"u}r Gravitationsphysik, Leibniz Universit{\"a}t Hannover and Max-Planck-Institut f{\"u}r Gravitationsphysik (Albert-Einstein-Institut), Callinstra{\ss}e 38, 30167 Hannover, Germany}
\address{$^{9}$Institute for High-Energy Physics, University of Amsterdam, Science Park 904, 1098 XH Amsterdam, Netherlands}
\address{$^{10}$Università di Padova, Dipartimento di Fisica e Astronomia, I-35131 Padova, Italy}
\date{\today}

\begin{abstract}
Frequency dependent squeezing is the main upgrade for achieving broadband quantum noise reduction in upcoming observation runs of gravitational wave detectors. The proper frequency dependence of the squeezed quadrature is obtained by reflecting squeezed vacuum from a Fabry-Perot filter cavity detuned by half of its linewidth. However, since the squeezed vacuum contains no classical amplitude, co-propagating auxiliary control beams are required to achieve the filter cavity's length, alignment, and incident beam pointing stability. In our frequency dependent squeezing experiment at the National Astronomical Observatory of Japan, we used a control beam at a harmonic of squeezed vacuum wavelength and found visible detuning variation related to the suspended mirrors angular drift. These variations can degrade interferometer quantum noise reduction. We investigated various mechanisms that can cause the filter cavity detuning variation. The detuning drift is found to be mitigated sufficiently by fixing the incident beam pointing and applying filter cavity automatic alignment. It was also found that there is an optimal position of the beam on the filter cavity mirror that helps to reduce the detuning fluctuations. Here we report a stabilized filter cavity detuning variation of less than 10$\,\mathrm{Hz}$ compared to the 113\,Hz cavity linewidth. Compared to previously published results [Phys. Rev. Lett. 124, 171101 (2020)], such detuning stability would be sufficient to make filter cavity detuning drift induced gravitational wave detector detection range fluctuation reduce from $11\%$ to within $2\%$.
\end{abstract}

\maketitle

\section{Introduction}
Advanced gravitational wave detectors, such as Advanced LIGO \cite{aasi2015advanced}, Advanced Virgo \cite{acernese2014advanced}, and KAGRA \cite{aso2013interferometer}, are designed to be quantum noise limited over most of their sensitivity spectrum.  During their third observation period, the phase quadrature squeezed vacuum was successfully utilized in LIGO and Virgo, enabled a 3\,$\mathrm{dB}$ quantum noise reduction above sub-100\,Hz  \cite{acernese2019increasing,tse2019quantum}. However, at the same time, the low frequency quantum noise originating from amplitude quadrature vacuum fluctuation was increased due to the Heisenberg uncertainty principle and approached the limit of the low frequency noise budget \cite{buikema2020sensitivity, acernese2020quantum,yu2020quantum}. 
Thus, a broadband quantum noise reduction becomes indispensable for further sensitivity improvement, and requires the use of frequency dependent squeezed vacuum \cite{kimble2001conversion}. Squeezed vacuum can obtain frequency dependence by reflection from a detuned optical cavity, called filter cavity. For advanced gravitational wave detectors operating with tuned signal-recycling cavity, one filter cavity having hundred Hertz linewidth and half-linewidth detuned is optimal \cite{purdue2002practical, khalili2010optimal, evans2013realistic, kwee2014decoherence}. Below 100\,Hz, phase quadrature squeezed vacuum will be rotated to the amplitude quadrature, while high frequency squeezed vacuum  is unchanged. Frequency dependent squeezed vacuum sources were developed in prototype \cite{zhao2020frequency,mcculler2020frequency} and will be used for the upcoming observation runs of Advanced Virgo+ \cite{flaminio2020status}, aLIGO+ \cite{barsotti2018a+} and KAGRA \cite{michimura2020prospects}. To operate an optical cavity, the laser frequency fluctuation and/or cavity mirrors' differential motion need to be controlled so that the cavity prompt reflected light can have a fixed phase difference with the light that enters, circulates and is leaked back from the cavity. When this phase difference is kept to be $\pi$, we say that the cavity is locked on resonance. But when this phase difference is held at an offset from $\pi$, we say that the cavity is locked with a detuning. In our experiment, the filter cavity length stability necessary for frequency dependent squeezing is achieved using an on-resonant 532\,nm (green) beam, which is a harmonic of the 1064\,nm (infrared) squeezed vacuum wavelength. The technique of controlling a laser beam's frequency using its own harmonic is referred to in this paper as bichromatic length control, and has been used for filter cavities \cite{oelker2016audio, zhao2020frequency}, optical parametric oscillators (OPOs) \cite{vahlbruch2016detection, stefszky2010investigation}, and arm length stabilization of gravitational wave detectors \cite{akutsu2020arm}. However, initially, using bichromatic length control yielded a filter cavity detuning stability of the order of 30\,Hz \cite{zhao2020frequency} while the final goal is to achieve 1\,Hz stability \cite{aritomi2020control}. 

In this paper, we study and identify the origin of the detuning drift and we demonstrate how to reduce it to a level compatible with the requirement of gravitational wave detection. Section \ref{sec:experiment} shows the experimental setup used to produce frequency dependent squeezing. Section \ref{sec:mechanism} outlines the mechanisms of sources of relative detuning drift for green and infrared beams. Section \ref{sec:verify} gives the results of measurements performed on the most prominent sources of relative detuning drift. Section \ref{sec:stability} shows the end result of improving the detuning stability of the filter cavity. Section \ref{sec:impact} discusses the implication of the improved results in the context of detection range of representative gravitational wave sources.

\section{Experimental setup}\label{sec:experiment}

\begin{figure}[t!]
    \includegraphics[scale=0.13]{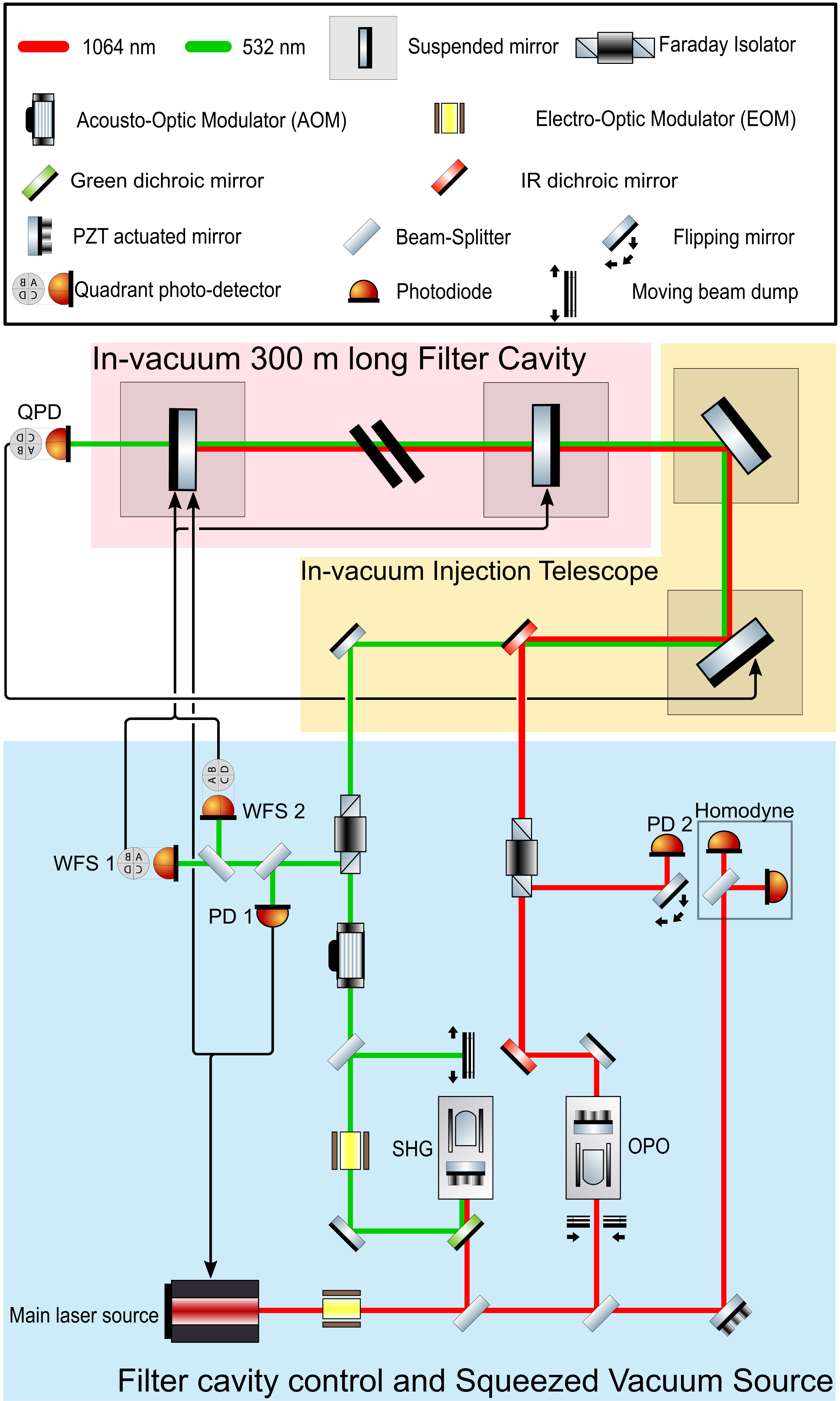}
    \caption{Simplified scheme of the experimental setup. The main laser beam is split into three paths. The first produces the green field at SHG. The second path is called bright alignment beam, which resonates inside OPO and goes to filter cavity. The third path is the local oscillator for the homodyne detector. The green beam is split into two paths: one controls the filter cavity and another pumps the OPO to generate squeezed vacuum. The bichromatic control signals are obtained from four detectors: PD 1 is used for the filter cavity length, WFS1 and WFS2 are used for the filter cavity alignment and a QPD in filter cavity transmission is used for incident beam pointing. This scheme indicates the beam paths used for characterization of filter cavity detuning at PD2. However, if two beam dumps and flipping mirror are changed, we can switch the system to generate and characterize frequency dependent squeezing. The beam path of squeezing is not shown but it is emitted from the OPO and will overlap with the bright alignment beam on the path to the filter cavity and back to the homodyne.}
\label{fig:AA_scheme}
\end{figure}

The experimental setup required to generate and characterize frequency dependent squeezing is introduced in our previous work \cite{zhao2020frequency}. In this paper, we focus on the detuning stabilization using newly implemented bichromatic control of filter cavity alignment and incident beam pointing. A simplified scheme of the experimental setup is shown in Fig.\,\ref{fig:AA_scheme}. Some important parameters of this filter cavity setup are summarized in Table.\,\ref{tab:FCpara}.

\begin{table}[t!]
\caption{\label{tab:FCpara}Summary of the filter cavity parameters. The values are nominal of measured parameters.}
\begin{ruledtabular}
\begin{tabular}{lcr}
Parameter & Symbol & Value \\
\hline
Length & L & \SI{300}{m}\\
Mirror radius & R & \SI{50}{mm}\\
Input mirror radius of curvature & $R_1$ & \SI{436.7}{m}\\
End mirror radius of curvature & $R_2$ & \SI{445.1}{m} \\
\hline
\multicolumn{3}{c}{\textsc{Parameters for 532} nm \textsc{beam}} \\
Input mirror transmissivity & $T_1$ & 0.7 \% \\
End mirror transmissivity & $T_2$ & 2.9 \% \\
Finesse & $\mathcal{F}$ & 172 \\
Beam radius at input mirror & $w_{1}$ & 7.26 mm \\
Beam radius at end mirror & $w_{2}$ & 7.35 mm \\
\hline
\multicolumn{3}{c}{\textsc{Parameters for 1064} nm \textsc{beam}} \\
Input mirror transmissivity & $T_1$ & 0.136 \% \\
End mirror transmissivity & $T_2$ & 3.9 ppm \\
Finesse & $\mathcal{F}$ & 4425 \\
Beam radius at input mirror & $w_{1}$ & 10.26 mm \\
Beam radius at end mirror & $w_{2}$ & 10.40 mm
\end{tabular}
\end{ruledtabular}
\end{table}

The main laser beam is split into three paths, one is used to pump the second harmonic generator (SHG), another is used as bright alignment beam, and the last one is used as local oscillator in homodyne detector. Additionally, the green beam is split into two paths: one is used for filter cavity bichromatic control and the other pumps OPO for squeezed vacuum generation. In the bichromatic control scheme, we use the green beam to control filter cavity length, alignment and incident beam pointing. The Pound-Drever-Hall technique \cite{black2001introduction} is used to extract the cavity length error signal at PD\,1 in reflection of the filter cavity. The cavity length correction signal is split into two frequency bands: the part below 10\,Hz is sent to the filter cavity end mirror position while the part above 10Hz is sent to the main laser frequency. Automatic alignment control uses two wavefront sensing quadrant photodiodes in the filter cavity reflection and acts on cavity mirrors' angle to keep the filter cavity axis aligned with the incident beam. (See Supplement Material \cite{Note1} first three sections for details of an optical cavity misalignment and the wavefront sensing automatic alignment implementation.) Incident beam pointing control is achieved using an error signal extracted from a quadrant photodiode in the filter cavity transmission and actuated on the angle of an in-vacuum suspended steering mirror in the filter cavity injection path. The green and infrared beams are overlapped by a dichroic mirror located before the pointing control steering mirror. Thus, alignment and pointing controls act on the co-propagating green and infrared beams. However, prior to the dichroic mirror, the green and infrared optics can move, causing a relative drift of alignment. A misalignment of few percent (in terms of power coupled with higher order modes) arises after several days.

An acousto-optic modulator located on the green beam going to filter cavity shifts the frequency of the green beam by the amount of a driving signal. In this way, after the green beam is locked on resonance for filter cavity, infrared detuning can be precisely adjusted by changing driving signal frequency.

The bright alignment beam resonates in the OPO, and the transmitted component has the same shape and propagation path as the squeezed vacuum. Meanwhile, bright alignment has phase modulated sidebands and is demodulated after its reflection from filter cavity. The acquired Pound-Drever-Hall signal allows us to investigate filter cavity detuning drift.

When beam dumps and flipping mirrors are changed in Fig.\,\ref{fig:AA_scheme}, squeezed vacuum is generated from OPO and obtains frequency dependent quadrature rotation after reflection from the detuned filter cavity. The homodyne detector is used to characterize the frequency dependent squeezing.

\section{Filter cavity detuning drift mechanisms}\label{sec:mechanism}

In bichromatic controls, in order to have a stable detuning for infrared, the green beam frequency change must be a factor two of the infrared beam frequency change. In addition, the green beam must be kept on resonance. However, many effects are involved in the experiment to change the filter cavity resonant condition, which are introduced in this section.

\subsection{Filter cavity length/laser frequency drift}\label{sec:mechanism_a}

An optical cavity needs to be locked to provide a stable amplitude/phase/mode-shape filtering for its incident light. A suspended cavity lock will be acquired if the control system can be engaged fast enough when a resonance is crossed, while it will be lost if a disturbance happens so that the control system cannot provide large enough correction \cite{barsotti2006control}. However, in the filter cavity bichromatic control, even during lock, there is residual drift of the cavity length and laser frequency that can influence the filter cavity detuning witnessed by the infrared beam. In addition to in-lock detuning drift, shift of the filter cavity detuning can also occur when the filter cavity loses lock. In the case of lock loss and re-acquisition, we find that only cavity length drift has contribution to the detuning variation. 

\textit{Detuning drift when the cavity is kept locked.}---During lock, the control system is working to compensate for residual filter cavity length drift $\delta \mathrm{L}$ by changing the laser frequency to hold resonance. There is a corresponding laser frequency shift of:
\begin{equation}
\label{eqn:l_change}
    \delta \mathrm{f} = \mathrm{f}\frac{\delta \mathrm{L}}{\mathrm{L}},
\end{equation}
where f is the laser frequency and L is the filter cavity length. The green beam frequency change is thus:
\begin{eqnarray}
\label{eqn:f_change}
\delta \mathrm{f}_{\mathrm{green}} = \frac{\mathrm{c}}{\mathrm{L}(\lambda_{\mathrm{green}}+\delta_{\mathrm{AOM}})}\delta \mathrm{L}.
\end{eqnarray}
The green laser frequency before entering the filter cavity is $\mathrm{f}_{\mathrm{green}} = \mathrm{c}/(\lambda_{\mathrm{green}}+\delta_{\mathrm{AOM}})$, where $\delta_{\mathrm{AOM}}$ is the green beam wavelength ($\lambda_{\mathrm{green}}$) change induced by AOM, c is the speed of light. Since the green beam is generated as a second harmonic of light from main laser, the infrared main laser frequency change is half of Eq.\,\ref{eqn:f_change}.
However, as seen in Fig.\,\ref{fig:AA_scheme}, when the setup is configured to generate squeezing, the OPO is pumped by green light sampled prior to the AOM. As such, the frequency of squeezed vacuum is half that of green prior to entering the AOM, which is the same as the frequency of main laser and thus the bright alignment beam. This is the reason why monitoring the bright alignment beam can predict the frequency behavior of squeezed vacuum. However, since the frequency of squeezed vacuum doesn't involve AOM, a frequency change of half of Eq.\,\ref{eqn:f_change} without $\delta \mathrm{AOM}$ is required to maintain the resonant condition for squeezed vacuum. There is a resulting change in the relative detuning between green and infrared beams as a consequence of the length correction:
\begin{equation}
\label{eqn:change}
    \Delta_{\mathrm{d}} = \frac{\mathrm{c}\cdot \delta_{\mathrm{AOM}}}{2\mathrm{L}\lambda_{\mathrm{green}}(\lambda_{\mathrm{green}}+\delta_{\mathrm{AOM}})}\delta \mathrm{L}.
\end{equation}
Since L = 300\,m and $\delta_{\mathrm{AOM}} = 2.1\times 10^{-13}$\,m (from AOM frequency shift $\mathrm{f}_{\mathrm{AOM}}$ = 110\,MHz), we can find a relation between detuning change and correction signals to filter cavity length and main laser frequency (details are provided in Supplement Material \cite{Note1} section four),
\begin{eqnarray}
\label{eqn:d_change}
    \Delta_{\mathrm{d}} [\mathrm{Hz}] &\simeq& 1.83\times10^5 [\mathrm{Hz/m}] \cdot\delta \mathrm{L} [\mathrm{m}], \\
    \label{eqn:d_change2}
    \Delta_{\mathrm{d}} [\mathrm{Hz}] &\simeq& 1.95\times10^{-7} \cdot\delta \mathrm{f} [\mathrm{Hz}].
\end{eqnarray}

\textit{Detuning shift after the cavity lock is lost and re-acquired.}---After lock loss and re-lock, the cavity resonant frequency may shift by an arbitrary number of free spectral ranges ($\mathrm{FSR} = \mathrm{c/2L} \simeq 500\,\mathrm{kHz}$). 

In the case of laser frequency shift during re-lock, the green beam frequency can change by $\mathrm{N'}\times \mathrm{FSR}\,(\mathrm{N'} \in \mathbb{Z})$. We find the infrared beam has no detuning change since the infrared beam frequency changes $\mathrm{N}'/2\times\mathrm{FSR}$. Note that we need to operate filter cavity with $\mathrm{N}'/2 \in \mathbb{Z}$. (Some equation derivations of this paragraph are provided in Supplement Material \cite{Note1} section four)

In the case of a cavity length change $\Delta \mathrm{L}$ during re-lock, FSR changes to $\mathrm{FSR}' = \mathrm{c/2(L+\Delta L)}$. This causes a detuning variation (details are provided in Supplement Material \cite{Note1} section four) :
\begin{eqnarray}
\label{eqn:d_change3}
\Delta_{\mathrm{d}} [\mathrm{Hz}] &=& (\mathrm{f}_{\mathrm{AOM}} \mod \mathrm{FSR}')\times\mathrm{FSR}'/2 \nonumber \\
&\simeq& 1.83\times10^5 [\mathrm{Hz/m}] \cdot\Delta \mathrm{L} [\mathrm{m}].
\end{eqnarray}
While this equation appears similar to Eq.\,\ref{eqn:d_change}, the cause of detuning shift is different. Eq.\,\ref{eqn:d_change} is derived when the cavity is locked at a specific resonance, but here the cavity is reasonably assumed to cross several FSRs after unlock.

In this experiment, sources of drift in laser frequency and cavity length are listed as follows.
\begin{itemize}
    \item Tides introduce a daily strain change of about $3\times10^{-8}$\cite{araya2017design}, corresponding to $9\mu$m length and 1.6\,Hz detuning drift.
    \item Ground motion introduces a weekly strain change of about $3\times10^{-9}$ \cite{sagiya2004decade}. This corresponds to $0.9\,\mu$m length and 0.2\,Hz detuning drift.
    \item A daily change of temperature introduces an estimated 60\,MHz of frequency drift at the main laser, corresponding to 11.7$\,\mathrm{Hz}$ detuning drift.
    \item A cavity alignment change can introduce a cavity length change (See Supplement Material \cite{Note1} section one for details of angular to length motion coupling). Assuming a 50$\,\mathrm{\mu rad}$ angle shift for the cavity mirrors, the maximum cavity length change is 2.2$\,\mathrm{\mu m}$. The corresponding detuning shift is 0.4$\,\mathrm{Hz}$.
\end{itemize}

From the consideration above, we see that the main laser frequency drift causes the most significant detuning shift while filter cavity is locked. For gravitational wave detectors, this issue can be mitigated by phase locking the squeezer main laser to the interferometer main laser, whose RMS frequency noise can be stabilized with a reference cavity \cite{kwee2012stabilized}. After acquiring interferometer arm lock, the frequency noise will be reduced at high frequency by transferring the lock of the main laser frequency to the common arm motion of the interferometer. But this doesn't contribute to a long term RMS frequency noise. Possibly, at this stage, the main laser frequency is still locked to a reference cavity at low frequency. This can achieve a RMS frequency noise reduction from ~60\,MHz to ~0.8\,MHz\cite{kwee2012stabilized}. However, such reference cavity was not used for experiment described in this paper.

\subsection{AOM driving signal frequency drift}

The AOM should provide a fixed frequency shift for green beam, but AOM driving signal source can have frequency drift. The infrared beam will acquire frequency drift of half of that imposed on the green beam. Using an oven-controlled crystal oscillator \cite{wenzel} as signal source, it is possible to keep detuning drift below 0.4\,Hz per week.

A minimum green frequency shift can be around 100\,Hz, so we plan to use two AOMs in sequence and impose opposite frequency shift. This will reduce $\delta_{\mathrm{AOM}}$ in Eq.\ref{eqn:change} and mitigate detuning drift issue. In addition, if two AOMs are driven by the same signal board, the AOM driving frequency drift can be cancelled. We also notice that the frequency drift issue can be mitigated by phase locking the AOM driving signal to a GPS.

\subsection{Different filter cavity length shift for green and infrared}

Normally, since the green beam overlaps with the infrared beam inside filter cavity, they should sense the same cavity length. However, when the green and infrared beams sense a different filter cavity length shift $\delta \mathrm{L_{IR}}$ and $\delta \mathrm{L_{GR}}$ respectively, the control system brings the green beam back to resonance, while the infrared detuning will shift by:
\begin{eqnarray}
\label{eq:diff_L}
    \Delta_{\mathrm{d}} &=& \mathrm{FSR}\left(\frac{\delta \mathrm{L_{infrared}}}{\lambda_{\mathrm{infrared}}/2}-\frac{\delta \mathrm{L_{green}}}{(\lambda_{\mathrm{green}}+\delta_{\mathrm{AOM}})}\right) \nonumber \\
    &\simeq& 9.4\times 10^{11} [\mathrm{Hz/m}]\cdot (\delta \mathrm{L_{IR}}-\delta \mathrm{L_{GR}}) [\mathrm{m}].
\end{eqnarray}
Approximation is made considering typical values in Eq.\,\ref{eq:diff_L}. 

\begin{figure}[t!]
    \includegraphics[width=0.45\textwidth]{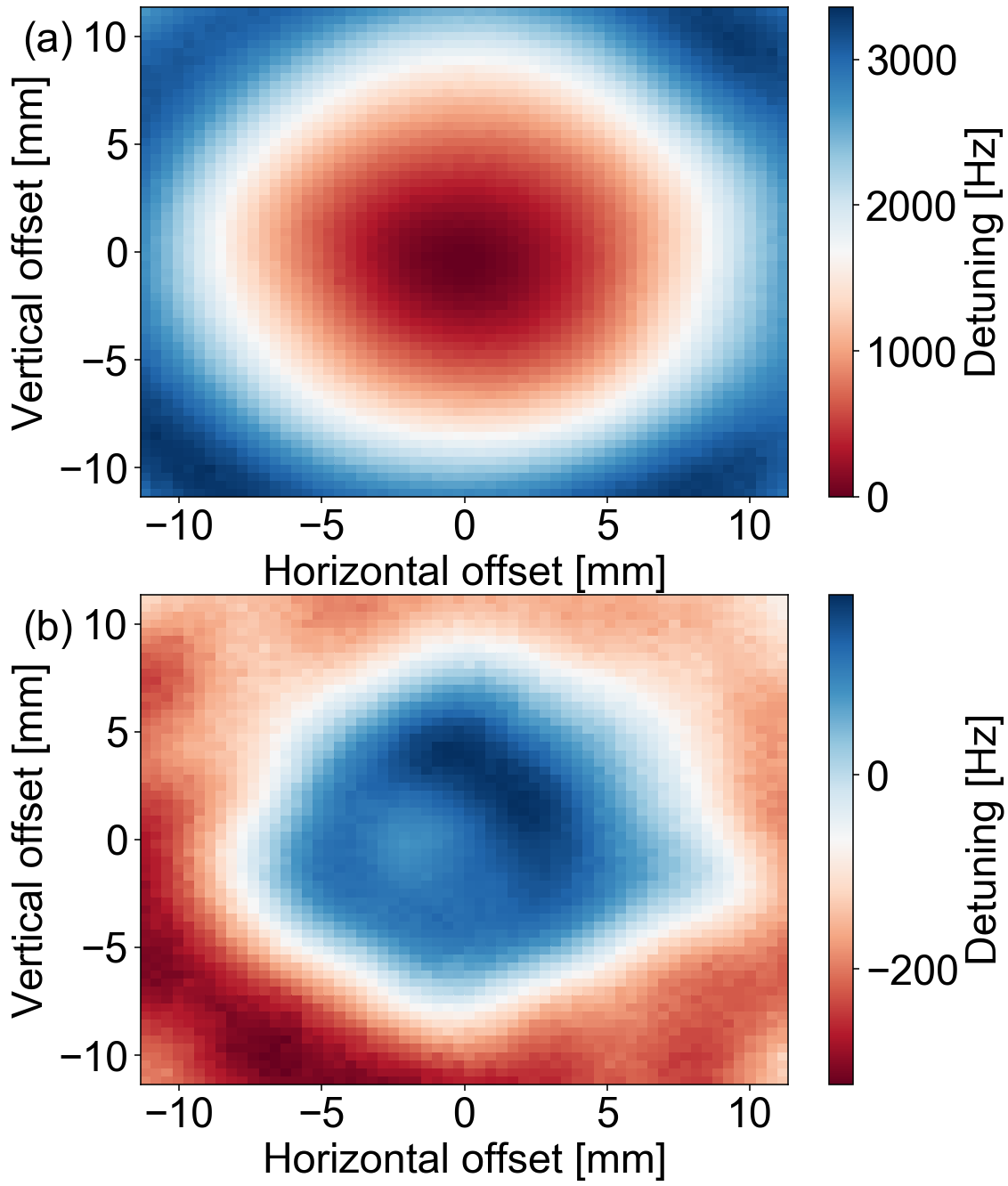}
    \caption{Simulated detuning variation between green and infrared beams caused by different beam position on the filter cavity end mirror. (a) Relative detuning shift as a result of green and infrared beams sensing different radius of curvature on different positions of the filter cavity end mirror. (b) Relative detuning variation as a result of green and infrared beams sensing different averaged mirror height on different position of filter cavity end mirror. The zero position is the center of end mirror. The detuning is adjusted in the simulation code to make it to be zero at the center of end mirror. }
\label{fig:map}
\end{figure}

The following mechanisms can cause the green and infrared beams to sense different filter cavity length:
\begin{itemize}
\item Temperature drift in the filter cavity mirror's dielectric coatings induces changes in their thickness and refractive index \cite{evans2008thermo, yam2015multimaterial}. The detuning change caused by this phenomenon is found to be much smaller than 1Hz during one day for our experiment \cite{Eisenmann2020phd}.

\item The wavefront error given by coating imperfections was found to have a dependence on wavelength \cite{sassolas2018high}. Additionally, the wavefront error is inhomogeneous over the mirror surface. This effect, when combined with alignment drift, causes a differing change in optical path length for the green and infrared beams. However, the wavefront error measurement of our filter cavity mirrors has only been performed at 633$\,\mathrm{nm}$ \cite{capocasa2016estimation}, so we cannot estimate this effect on filter cavity detuning stability. 

\item Similarly to the above, the wavefront error also has a dependence on beam size, which is different between the green and infrared beams as shown in Table.\,\ref{tab:FCpara}. A simulation was performed where different Gaussian weightings are required to evaluate how the mirror wavefront deviates from a perfect sphere. From this simulation, the results of which are shown in Fig.\,\ref{fig:map}, we find a detuning variation of few kHz could happen. This is because a different RoC and averaged mirror height will be sensed by green and infrared beams after a cavity alignment change. More details of this simulation are provided in Supplement Material \cite{Note1} section five. However, due to the lack of the wavefront error measurements at 532$\,\mathrm{nm}$ and 1064$\,\mathrm{nm}$, the simulation is not completely accurate.
\end{itemize}

\subsection{Residual amplitude modulation of sidebands}

The Pound-Drever-Hall locking technique utilizes the beat signal between the carrier and phase modulated sidebands. However, the error signal has fluctuations due to residual amplitude modulation (RAM) that arises from various imperfections in the phase modulation process \cite{whittaker1985residual, wong1985servo}. In the past, RAM was found to produce noise in gravitational wave detectors \cite{kokeyama2014residual}, and cause drift in the level of generated squeezing \cite{li2016reduction}.

In this experiment, RAM is introduced at the green EOM, which causes a locking point error for the green beam and a corresponding detuning shift of infrared. In addition, RAM from the infrared EOM is present when bright alignment beam is used to characterise detuning. 

\section{Detuning drift sources identification}\label{sec:verify}

In this section, we report on the experimental identification of relative detuning sources of the filter cavity for green and infrared beams.

\begin{figure}[t!]
    \includegraphics[width=0.45\textwidth]{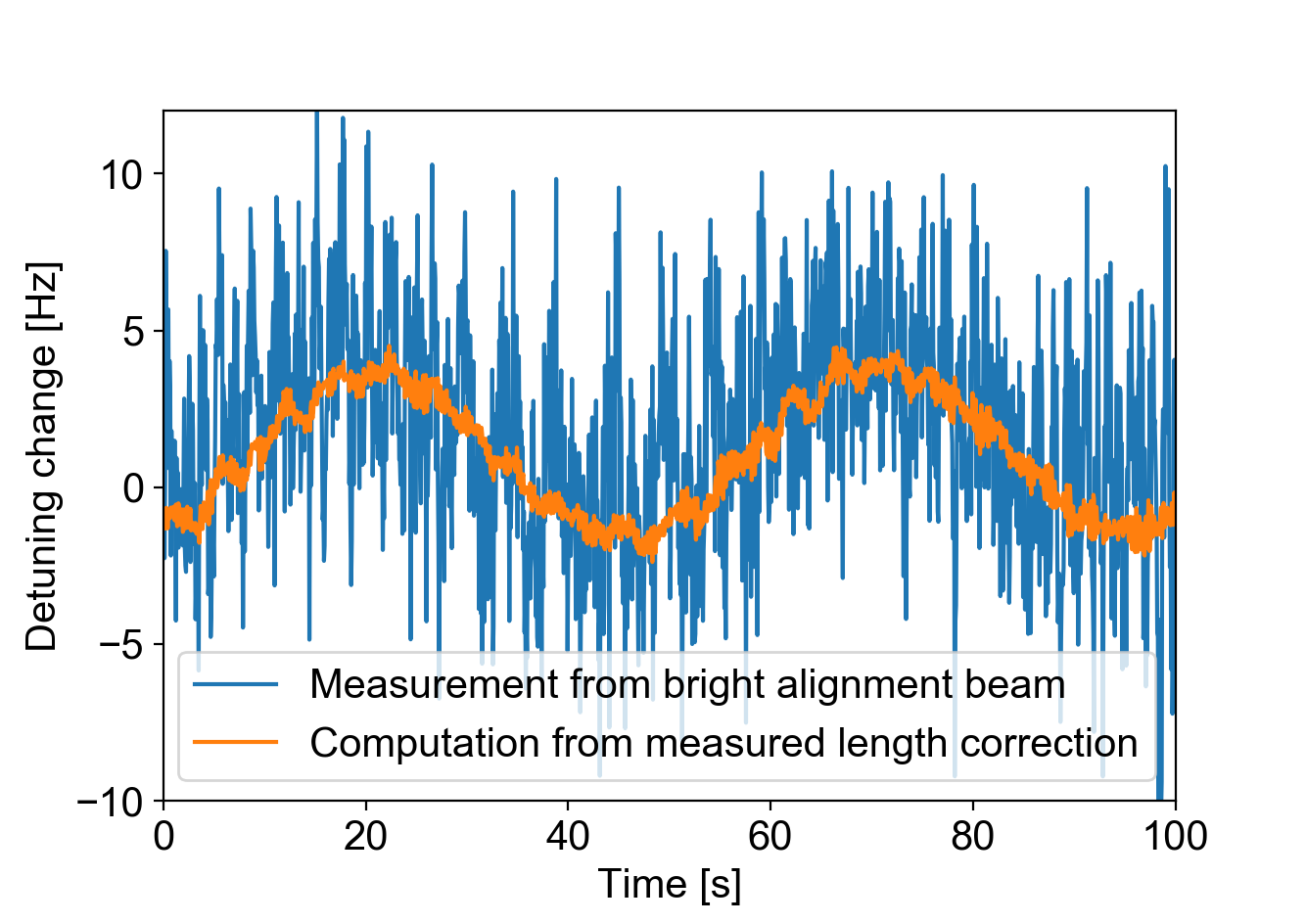}
    \caption{Detuning change when the main laser frequency is sinusoidally varied. The resulting length correction signal is measured over a period of 100s and converted to a detuning variation as per Eq.\,\ref{eqn:d_change} (orange). This is found to overlap with the detuning change measured by the bright alignment beam.}
\label{fig:comb1}
\end{figure}

\textit{Testing detuning change with laser frequency.}---In order to experimentally test Eq.\,\ref{eqn:d_change}, we introduce a sinusoidal main laser frequency shift and check detuning. The main laser frequency shift is introduced by changing the main laser crystal temperature. The control loop responses to the main laser frequency shift and corrects on the filter cavity end mirror position to keep the green beam locked on resonance. This correction signal $\delta \mathrm{L}$ is acquired and taken into Eq.\,\ref{eqn:d_change} to obtain a detuning variation $\Delta_{\mathrm{d}}$, which shows as the orange curve in Fig.\,\ref{fig:comb1}. On the other hand, the filter cavity reflected bright alignment beam contains RF sidebands. A demodulation of this field at RF sideband frequency gives a PDH signal, which is calibrated in Hz as the blue curve in Fig.\,\ref{fig:comb1}. These two curves in Fig.\,\ref{fig:comb1} match well, which proves that Eq.\,\ref{eqn:d_change} describes well a detuning change caused by a length correction signal of the cavity length control loop.

\begin{figure}[t!]
    \includegraphics[width=0.45\textwidth]{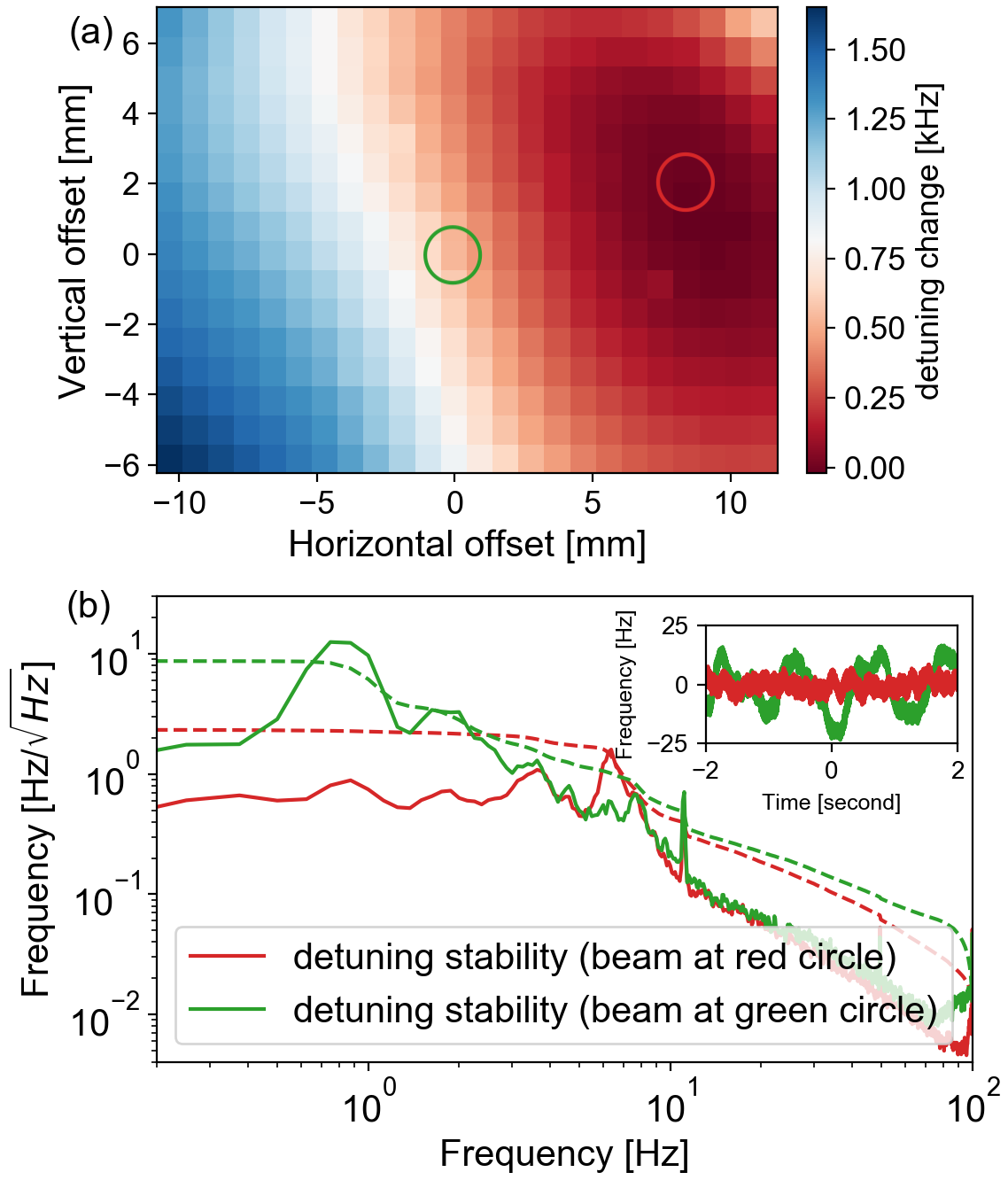}
    \caption{Detuning drift dependence on alignment. (a) Detuning change with different beam positions on filter cavity end mirror. The dark red area (indicated by a red circle) is the region where we should place the beam for the best detuning stability. We set the detuning at this point to be zero since it is the nominal operation point for the pointing control loop. (b) Detuning stability for two different beam positions on end mirror. Detuning spectrum (green) with internal cavity beam hitting on the end mirror center (indicated by a green circle) has integrated RMS (green dashed) value of 8.7\,Hz. Detuning spectrum (red) with internal cavity beam hitting on the end mirror stable region (indicated by a red circle) has integrated RMS (red dashed) value of 2.3\,Hz. The inset shows detuning fluctuation as a function of time corresponding to the respective colors.}
\label{fig:comb2}
\end{figure}

\textit{Testing detuning change with alignment.}---Alignment-dependent detuning shift is measured by offsetting pointing control to scan a 13$\times$23\,mm region on the filter cavity end mirror. At each position on the end mirror, we adjusted the AOM driving frequency to make the bright alignment beam resonant together with the green beam. By taking the AOM adjusted frequency relative to its nominal value, we get a resulting detuning shift mirror map as shown in Fig.\,\ref{fig:comb2} (a). A maximum detuning shift of 1.6\,kHz was found. Similarly, we check detuning dependence on the input mirror beam position, where it was found that the spatial dependence of detuning was about 15 times less significant. 

Fig.\,\ref{fig:comb2} (a) shows that intra-cavity beam jittering around the dark red region (indicated by a red circle) should introduce less detuning fluctuation compared to the mirror center (indicated by a green circle). This is because the beam angular motion coupling to length variation has a smaller gradient at the position indicated by the red circle in Fig.\,\ref{fig:comb2} (a). We moved the beam position from the green circle position to the red circle position on the end mirror, then measured the detuning spectrum from the bright alignment beam. It was seen that the RMS detuning stability was improved by a factor of 4, as shown in Fig.\,\ref{fig:comb2} (b).

\textit{Testing detuning change due to residual amplitude modulation}---To measure the effect of RAM, we lock the filter cavity such that the bright alignment beam is anti-resonant. In this condition, the Pound-Drever-Hall error signal taken at the reflection of filter cavity is insensitive to phase change, and therefore only RAM can be seen in the error signal fluctuations around zero. We find that RAM can introduce detuning shift of around 4\,Hz within a period of tens of minutes. In the future, we plan to use wedged crystal EOM to reduce RAM \cite{li2016reduction}.
    
\section{Filter cavity detuning stabilization} \label{sec:stability}

After analyzing various detuning drift mechanisms, we find the most significant source is from the alignment drift and mirror jittering as shown in Fig.\,\ref{fig:comb2}. We used alignment and pointing controls to keep the beam on the stable region of the end mirror, then measured long term detuning stability and frequency dependent squeezing. 

\begin{figure}[htbp]
\begin{center}
    \includegraphics[width=0.45\textwidth]{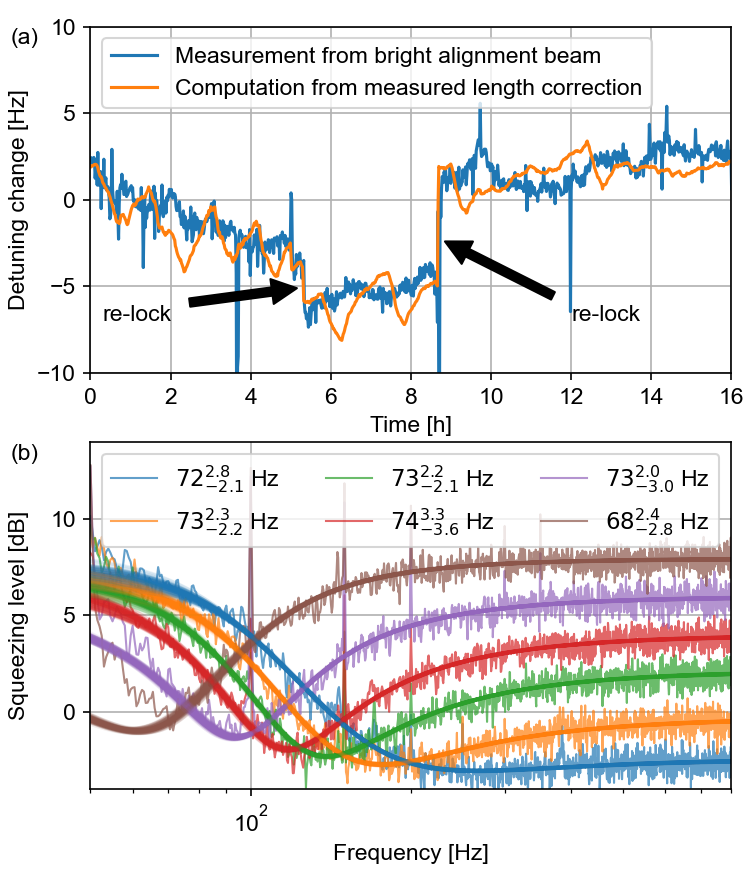}
    \caption{Filter cavity detuning drift characterization under bichromatic controls. (a) Filter cavity detuning change monitored with bright alignment beam. Measured detuning change (blue) overlaps with the implied detuning change calculated from the measured length correction signal and Eq.\,\ref{eqn:d_change} (orange). (b) Frequency dependent squeezing measurements using different homodyne angle under bichromatic controls. Each measurement (noisy curve) is averaged 100 times. For each measurement, we randomly plot 100 curves (smooth curves) found by `emcee' fit. The legend shows fit results of filter cavity detuning. At low frequency, backscattering from the motion of the suspended mirrors dominates over squeezing, which is the main reason for the measurement deviation from the fit results below $\sim$70\,Hz.}
\label{fig:long}
\end{center}
\end{figure}

Long term detuning stability was measured using the bright alignment beam while the filter cavity was locked with bichromatic controls for 16 hours. We obtained the filter cavity detuning change with the same method as used for Fig.\,\ref{fig:comb1}, described in section\,\ref{sec:verify}. The measured detuning is shown by the blue curve in Fig.\,\ref{fig:long} (a) and is seen to have a variation of 8.05\,Hz within 68\% percentile. The inferred detuning change from the length correction and Eq.\,\ref{eqn:d_change} is shown by the orange curve. In this plot, the overlap of blue and orange curves indicates that the detuning change mainly comes from the length correction signal caused by laser frequency drift, while their discrepancy is attributed to the RAM effect. 

There are two re-locks in Fig.\,\ref{fig:long} (a) causing the cavity length and detuning jump. In the first re-lock, we noticed that it happened suddenly. Approximately 10\,$\mu$m displacement happened for the controlled mirror, which is a reasonable mirror motion within one pendulum period. But in the second re-lock, we left time for the mirror pendulum motion damping, which released an accumulated mirror displacement of approximately 30\,$\mu$m and made the detuning go back to zero. As calculated in section \ref{sec:mechanism_a} and observed in Fig.\,\ref{fig:long} (a), the detuning variation after cavity re-lock is only related to the cavity length change. Correspondingly, although a substantial laser frequency drift happened, it didn't affect detuning after un-lock.

The detuning stability is also checked by characterizing frequency dependent squeezing as shown in Fig.\,\ref{fig:long} (b) using different homodyne angles. The whole measurement took one hour. We use the model of frequency dependent squeezing of Kwee et al.\,\cite{kwee2014decoherence} to fit the measured data and extract the filter cavity detuning as one of the fit parameters, along with homodyne angle, propagation optical losses, and generated squeezing level. The fit is performed using `emcee' \cite{foreman2013emcee}, further details of which are provided in Supplemental Material \cite{Note1} section six. The fit results for six different homodyne angles are shown in the legend of Fig.\,\ref{fig:long} (b). A detuning fluctuation of $6.3\,\mathrm{Hz}$ is found by aggregating all fit results and checking the 68\% percentile. The fluctuation of $6.3\,\mathrm{Hz}$ is slightly less than that measured over the course of 16 hours using the bright alignment beam.

\section{Impact of improved filter cavity detuning stability on gravitational wave detection range} \label{sec:impact}

Frequency dependent squeezing requires the filter cavity to have a specific and stable detuning to optimize quantum noise reduction. To understand how the detuning drift impacts gravitational wave detector quantum noise, we make a simulation based on the frequency dependent squeezing model of Kwee et al. \cite{kwee2014decoherence} and the noise budget of the KAGRA gravitational wave detector \cite{kagra_s}. The result of frequency dependent squeezing with improved detuning stability in KAGRA is shown in Fig.\,\ref{fig:det_range}. More details about this simulation are provided in Supplement Material \cite{Note1} section seven. Fig.\,\ref{fig:det_range} (a) shows that a detuning change of $\pm 15$\,Hz can degrade the sensitivity such that there is no improvement from frequency dependent squeezing below 75\,Hz. However, reducing the detuning drift to $\pm 5$\,Hz makes the influence on detector sensitivity almost negligible. We characterize the improvement of KAGRA scientific output in terms of two relevant sources: 1)binary black hole (BBH) mergers of mass 50\,$\mathrm{M}_\odot$ as a representative of compact binary coalescences that occur at low frequency, 2) binary neutron star (BNS) detection. Their corresponding detection ranges are calculated using the inspiral-range package \cite{inspiral}. This low frequency stabilization of detector sensitivity would benefit the early warning of binary neutron star mergers \cite{cannon2012, low2019, yu2021, chu2022} and the detection of intermediate-mass black holes \cite{mandel2008rates, graff2015missing, veitch2015measuring, gw190521}. If filter cavity detuning stability is $\pm 15$\,Hz, the introduction of frequency dependent squeezing increases the BNS range by 22.0\% to 35.2\%, while the 50\,$\mathrm{M}_\odot$ BBH range by 5.4\% to 17.8\%. The insets of Fig.\,\ref{fig:det_range} (b) show that by reducing the detuning drift to $\pm$ 5\,Hz, the BNS range has a minimum boost of 33.4\%, while the BBH range has a minimum boost of 16.0\%. This corresponds to an overall range fluctuation reduction from 10.6\% to 1.5\%.

\begin{figure}[t!]
    \includegraphics[width=0.45\textwidth]{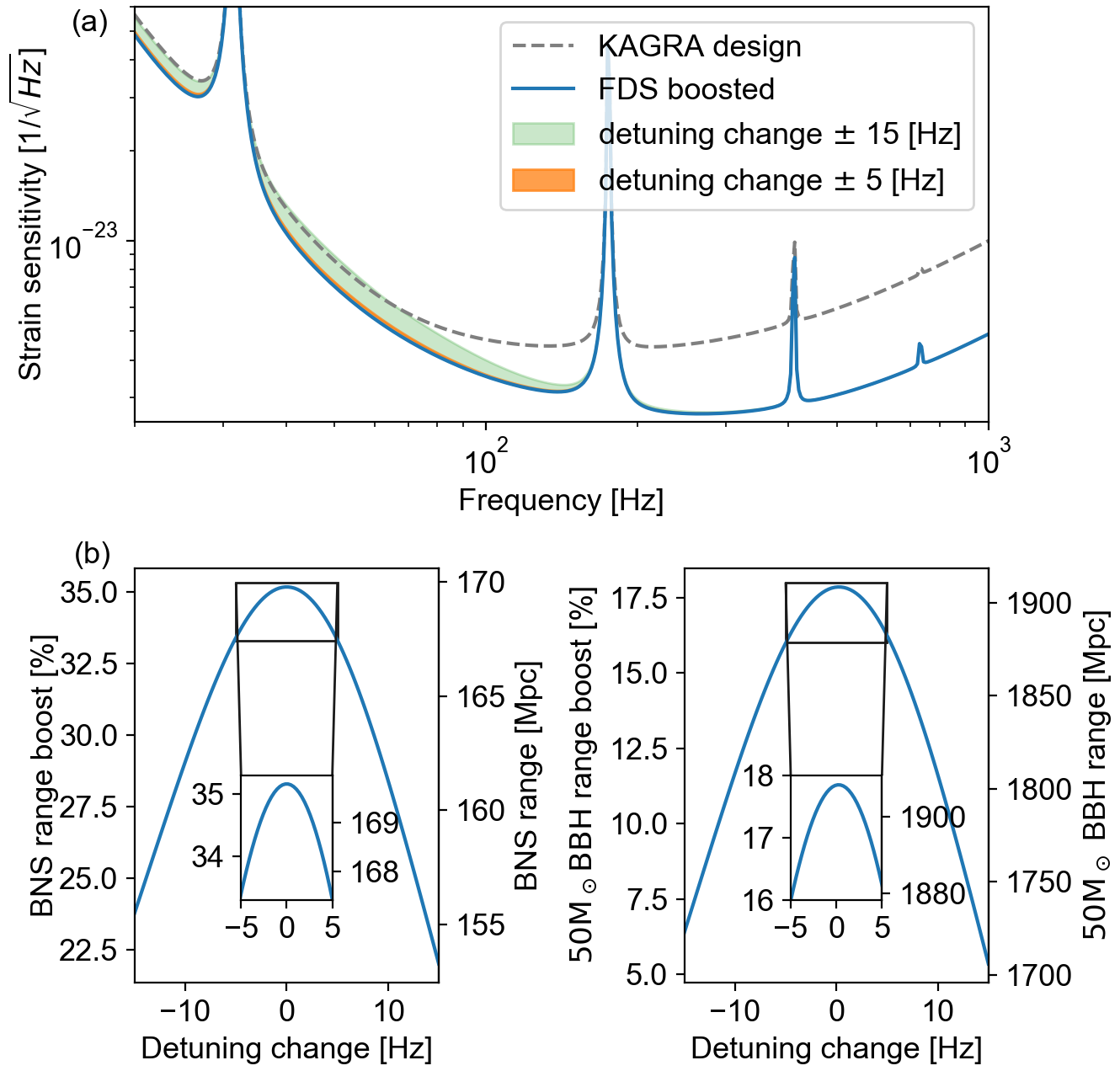}
    \caption{Influence of filter cavity detuning drift on KAGRA scientific outcomes. (a) Assuming the filter cavity detuning has no fluctuation, we estimate that the KAGRA designed broadband resonant sideband extraction mode noise (grey dashed) can be reduced across a broadband of frequencies (blue). The effect of previously observed filter cavity detuning fluctuation $\pm 15$\,Hz is shown in the green shaded area. The effect of the improved detuning fluctuation $\pm 5$\,Hz is shown by the orange shaded area. (b) Detector range versus filter cavity detuning shift. The insets show zoom-in of $\pm 5$\,Hz detuning shift. }
\label{fig:det_range}
\end{figure}

\section{Conclusion and outlook}

The stability of frequency dependent squeezed vacuum is crucial for reducing quantum noise in upcoming observing runs of gravitational wave detectors. We have developed bichromatic automatic filter cavity alignment and incident beam pointing controls for the infrared squeezed field using an overlapping harmonic green beam. We have investigated mechanisms that cause infrared detuning variation of the filter cavity, and found that the most significant contribution comes from alignment drift of the filter cavity mirrors combined with the beam position on the end mirror. The mirror jittering induced detuning fluctuation was reduced by positioning the intra-cavity beam at a stable region on the end mirror, reducing the RMS detuning fluctuation by a factor of four. After that, we have achieved frequency dependent squeezing with a detuning stability smaller than 10 Hz, an improvement of a factor of three from previously published results \cite{zhao2020frequency}. Such FC detuning stability is crucial to fully take advantage of frequency dependent squeezing and avoid degrading the detector range.

Further improvements in the detuning stability will come from mitigating the large frequency shifts that occur in the length correction loop, for example, by using two AOMs or stabilizing main laser frequency. We should use wedged crystal EOMs to reduce residual amplitude modulation in Pound-Drever-Hall error signals. Although detuning drift caused by different mirror wavefront errors for green and infrared beams can be mitigated, their origin is not fully understood and should be further investigated.

For gravitational wave detectors, the squeezer laser is phase locked to the stable interferometer main laser, so we don't expect detuning drift caused by laser frequency drift while the cavity is locked. Residual amplitude modulation induced detuning drift should be small since wedged crystal EOMs are usually employed in gravitational wave detectors. The filter cavity length change maybe an issue since ground motion and mirror suspension stability are different for different gravitational wave detectors, but this can be reduced by reducing green beam frequency shift using two AOMs.

The filter cavity intra-cavity optical losses has dependence on alignment \cite{isogai2013loss, capocasa2018measurement}, which can be investigated with bichromatic controls more precisely in the future.

There are also filter cavity control schemes that use infrared sidebands which are mode matched and co-propagating with the squeezed vacuum. These have been dubbed as the coherent control filter cavity scheme \cite{aritomi2020control, aritomi2021} and resonant locking field scheme \cite{mcculler2020frequency}. These schemes offer the advantage of ensuring a proper sensing of the filter cavity length noise for the squeezed vacuum. However, due to the configuration of the filter cavity compared to gravitational wave detectors, bichromatic control allows for comparatively easier lock acquisition and larger sensing power. If bichromatic controls are improved to the point where they can give close to ideal alignment of the squeezed field, then it will greatly ease the commissioning and lock acquisition of a gravitational wave detector using frequency dependent squeezing and provide faster improvement of scientific outcomes.

\section*{Acknowledgements}
We thank J. Degallaix, X. Ding, and T. Liu for their contributions and discussions. We thank G. Hartmut and A. Jones for their comments on this work. We thank also Advanced Technology Center of NAOJ for the support. We acknowledge the help from members of KAGRA Collaboration, Virgo, and LIGO Collaboration. This work was supported by the JSPS Grant-in-Aid for Scientific Research (Grants No.\,15H02095, No.\,18H01235, and No.\,21H04476), the JSPS Core-to-Core Program, and the EU Horizon 2020 Research and Innovation Programme under the Marie Sklodowska-Curie Grant Agreement No. 734303. Y.\,Z. was supported  by the Graduate University of Advanced Studies, SOKENDAI, by the Japanese government MEXT scholarship, and by the ICRR Young Researcher's Fund. M.\,E. was supported by the JSPS Standard Postdoctoral Fellowship (20F20803). M.\,P. was supported by the JSPS Standard Postdoctoral Fellowship (20F20713). H.\,L. and H.\,V. were supported by the Deutsche Forschungsgemeinschaft (DFG, German Research Foundation) under Germany's Excellence Strategy - EXC 2123 QuantumFrontiers - 390837967. N.\,A. was supported by JSPS Grant-in-Aid for Scientific Research (Grant No.\,18H01224), JSPS Grant-in-Aid for Challenging Research (Exploratory) (Grant No.\,18K18763), and JST CREST (Grant No.\,JPMJCR1873).

\end{document}